\documentclass[journal]{IEEEtran}
\ifCLASSINFOpdf
\else
\usepackage[dvips]{graphicx}
\fi

\usepackage[caption=false]{caption}
\usepackage[font=footnotesize]{subfig}
\usepackage{fixltx2e}

\usepackage[nospace]{cite}

\newcommand {\chsq}   {\mbox{${\chi}^2$}}

\begin{document}
%
\title{Simulations of a Scintillator Compton Gamma Imager for Safety and Security}
%
%
%

\author{L.E.Sinclair,
        D.S.Hanna,
        A.M.L.MacLeod,
        and P.R.B.Saull
\thanks{L.E.Sinclair, corresponding author, is with the Geological Survey of Canada, Natural Resources Canada, Ottawa, Ontario, Canada.  Correspondence: laurel.sinclair@nrcan.gc.ca}
\thanks{D.S.Hanna and A.M.L.MacLeod are with the Physics Department, McGill University, Montreal, Quebec, Canada}
\thanks{P.R.B.Saull is with the Institute for National Measurement Standards, National Research Council, Ottawa, Ontario, Canada}
\thanks{Published in IEEE Transactions on Nuclear Science.
        Copyright may be transferred without notice, after which this version
        may no longer be accessible.}}

%
%

\markboth{Presented at SORMA West 2008, Berkeley, California, June~2008}%
{Shell \MakeLowercase{\textit{et al.}}: Bare Demo of IEEEtran.cls for Journals}
%



\maketitle

\begin{abstract}
We are designing an all-scintillator Compton gamma imager for use in security investigations and remediation actions involving radioactive threat material.
To satisfy requirements for a rugged and portable instrument, we have chosen solid scintillator for the active volumes of both the scatter and absorber detectors.
Using the BEAMnrc/EGSnrc Monte Carlo simulation package, we have constructed models using four different materials for the scatter detector: LaBr$_3$, NaI, CaF$_2$ and PVT.
We have compared the detector performances using angular resolution, efficiency, and image resolution.
We find that while PVT provides worse performance than that of the detectors based entirely
on inorganic scintillators, all of the materials investigated for the scatter detector have the potential to provide performance adequate for our purposes.
\end{abstract}


%
\IEEEpeerreviewmaketitle

\section{Introduction}
%
%
%
\IEEEPARstart{A}{} requirement for innovative detection technologies to assist investigators in intelligence gathering prior to or after a radiological or nuclear incident has been identified by Canada's Chemical, Biological, Radiological-Nuclear and Explosives Research and Technology Initiative (CRTI).
To address this need, we are designing a Compton gamma imager.
Our design goal is a compact instrument capable of localizing a 10~mCi point source of Cs-137 40~m away to within a few degrees, in a field of view of 
$\pm 45^{\circ}$ in both directions, in under a minute.

There are other groups investigating related imager designs, employing HPGe~\cite{HPGedudes}, Si~\cite{SIdudes1,SIdudes2,SIdudes3}, CZT~\cite{CZTguys}, or gaseous time projection chamber~\cite{gasguys} detector technologies.
These techniques generally provide superior energy- and ultimately image resolution on a per-event basis, to what can be achieved with scintillator materials.
On the other hand, scintillators provide the benefit of a cost-effective way to produce a high-efficiency detector in a form which can readily be made compact and rugged, for deployment to the field.
This portability constraint also introduces a need for an instrument which has low power consumption, and the cost-effectiveness of scintillator opens the possibility of deployment of more than one unit.

There is an interesting approach to hybrid Compton and coded-aperture imaging which also uses an all-scintillator design~\cite{hybridguys}.
Spare parts from an all-scintillator space-borne Compton telescope have been used to demonstrate a capability at ground level of identifying sources of radiation~\cite{Compton_telescope}.
There has also been a study indicating that an all-scintillator design could be promising for the detection of highly enriched uranium~\cite{HEUstudy}.

Here, we present design studies conducted using the BEAMnrc/EGSnrc simulation package~\cite{BEAMcite,EGScite}.
This study aims to determine whether our design goal is achievable, and whether some prospective scintillator materials can be ruled out.
This work will proceed toward the development of a prototype.

\section{Compton Imaging}

\begin{figure}[!t]
\centering
\includegraphics[width=2.5in]{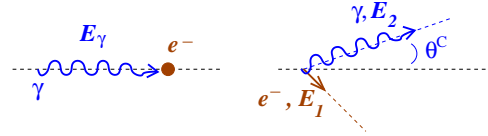}
\caption{Compton scatter diagram, a) initial-state and b) final-state.}
\label{fig_diag}
\end{figure}

The process of Compton scattering is illustrated in Figure~\ref{fig_diag}.  
An incoming photon of energy $E_{\gamma}$ scatters from an atomic electron, leading to a final state in which there is an outgoing electron of energy  
$E_1$ and an outgoing photon of energy $E_2$.

\begin{figure}[!t]
\centering
\includegraphics[width=2.in]{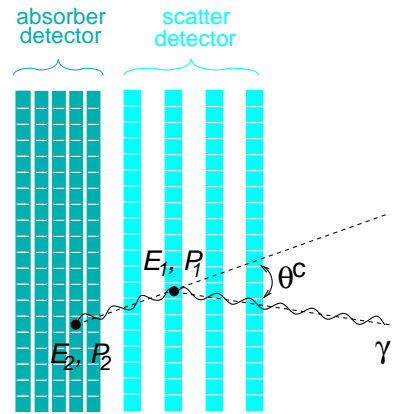}
\caption{Schematic diagram of a Compton gamma imager.}
\label{fig_schematic}
\end{figure}

A sketch of a Compton gamma imager is provided in Figure~\ref{fig_schematic}.
The energy $E_1$ is deposited at some location in a pixellated scatter detector.  The outgoing photon escapes the scatter detector to deposit its energy, $E_2$, at some location in a position-sensitive absorber detector.
The scattering angle between the initial and final state photons, $\theta^C$, can be determined from the two energy deposits, according to,
\begin{equation}
\cos \theta^C = 1 + m_0 c^2(\frac{1}{E_\gamma} - \frac{1}{E_2}),
\label{eqn1}
\end{equation}
where $E_\gamma = E_1 + E_2$ and $m_0 c^2$ is the electron rest energy.
Thus, the position of the source may be reconstructed to lie somewhere on a cone of opening angle $\theta^C$ with its axis along the line joining the positions of the two energy deposits, and its apex at the first energy deposit.  By back-projecting the cones from several events onto an image plane, an image may be reconstructed from the positions where the cones overlap.

\section{Detector Models}
Using the BEAMnrc/EGSnrc Monte Carlo simulation packages we have constructed models of Compton gamma imagers.  
The models consist of layers of scintillator 20~cm x 20~cm in cross section, with 1~cm thickness in the scatter detector and 0.8~cm thickness in the absorber detector.
Four different materials have been tested for the scatter detector, LaBr$_3$, NaI, CaF$_2$ and polyvinyltoluene-based plastic scintillator, hereinafter referred to as 
PVT\footnote{Densities for these materials are taken from the technical data provided by the commercial supplier Saint Gobain.
The PVT used in these simulations corresponds to Saint Gobain's general purpose scintillator BC-418.}.
In the following, the detector models will be referred to by the scatter detector material.
The number of scatter detector layers is dependent upon the material.

To determine the optimal thickness of the scatter detector for each material, we looked at the probability for an incoming gamma to Compton scatter or to undergo a photo-electric process, as a function of material thickness.  For this study, we have generated 10,000 events from a 662~keV 0.1~mm x 0.1~mm square parallel beam incident at the centre of the front face of a 50~cm x 50~cm x 50~cm cube of material.

\begin{figure*}[!t]
\centering
\includegraphics[width=5.in]{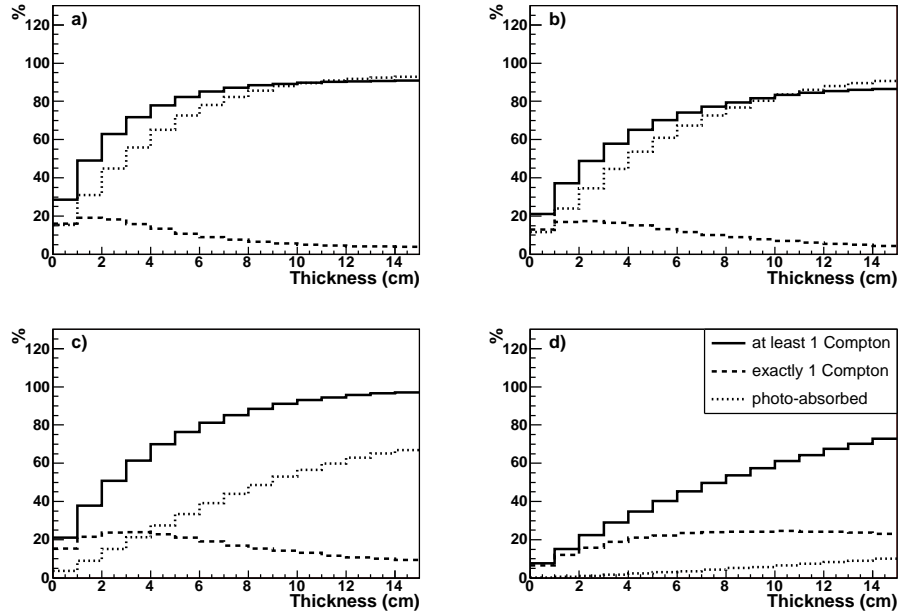}
\caption{Percentage of 662~keV events undergoing interaction as a function of slab thickness for a) LaBr$_3$, b) NaI, c) CaF$_2$ and d) PVT materials.  The solid histogram indicates the percentage of incoming gammas for which at least one Compton scatter will occur.  The dashed histogram shows the percentage which will have undergone exactly one Compton scatter.  The dotted histogram shows the percentage of events undergoing a photo-electric process.}
\label{fig_2}
\end{figure*}

The probability for an interaction to occur is presented in Figure~\ref{fig_2} as a function of depth within the slab.
For all materials, as the thickness of the material increases, the probability of at least one Compton scatter occuring, represented by the solid histogram in Figure~\ref{fig_2}, increases.
However, the probability of the photon being absorbed in a photo-electric process, represented by the dotted histogram, increases with thickness as well.
The probability of secondary Compton scatters also increases with material thickness (not shown).
This means that there is some thickness at which the probability of exactly one Compton scatter is maximized.  This probability is represented in Figure~\ref{fig_2} by the dashed histograms.

In the models discussed here, we have chosen the material thickness of the scatter detector according to the maximum of the dashed curve in Figure~\ref{fig_2}.
Thus the LaBr$_3$, NaI, CaF$_2$ and PVT scatter detectors feature two, three, four and eight, 1~cm-thick slabs of scintillator, respectively.  There is 2~cm of spacing between the front faces of successive layers of scintillator in the scatter detector.

All of the models feature an absorber detector consisting of five 0.8~cm layers of LaBr$_3$ with 1~cm spacing between the front faces of successive layers.

The angular resolution of a Compton imager will improve, the farther apart are the energy depositions $E_1$ and $E_2$.  At the same time, at least for these simple designs, the efficiency will worsen the farther apart are $E_1$ and $E_2$.  In order to reduce the differences in performance which are due to these geometrical differences between the different models, we have chosen to keep the distance between the centre of the scatter detector and the front face of the absorber detector the same for all models.  This distance is 9~cm, leaving air-gaps between the scatter detector and the absorber detector of various sizes for the different models, with the PVT detector having the smallest air-gap at 1~cm.

Readout devices for light collection placed between the layers will constitute dead material within the Compton imager.  To account for this effect we have added a 1~mm-thick layer of SiO$_2$ and a
10~$\mu$m-thick layer of Au after each scintillator layer in both the scatter and absorber
detectors, to represent arrays of silicon photomultipliers.

\section{Event Simulation}

To compare the performance of the detectors, we simulated a mono-energetic point-source situated 40~m from the centre of the front face of the models.
We investigated the energy dependence of the detector performance, choosing on-axis point sources of energy 300~keV, 500~keV, 1~MeV, 1.5~MeV and 2~MeV.
We also looked at the performance for a 662~keV source located on-axis and at angles between the detector axis and the line between the centre of the front face and the source position of 10$^{\circ}$, 20$^{\circ}$, 30$^{\circ}$ and 40$^{\circ}$.

We investigated three different sources of image degradation: initial-state electron physics effects, energy resolution effects, and the effect of finite detector segmentation.
\begin{itemize}
   \item In EGSnrc, binding effects and Doppler broadening are treated according to the relativistic impulse approximation~\cite{Ribberfors}.  These effects are controlled by an input parameter and may easily be turned on or off.
  \item Smearing of the energy measurement in the scintillator and readout was applied to energy deposits in the NaI scatter detector using energy resolutions determined by experiment~\cite{NaIres}.  
For all other materials, the energy deposits $E_i$ were smeared by a Gaussian distribution about the true energy deposited, of width $C \sqrt{E_i}$, where the constant $C$ was determined for LaBr$_3$, CaF$_2$ and PVT by the constraint that the FWHM resolution at 662~keV should come to 2.9\%, 10\% and 14\% 
respectively\footnote{2.9\% is a typical FWHM energy resolution at 662~keV for LaBr$_3$ quoted by suppliers.  
Note that suppliers do not quote typical energy resolutions at 662~keV for CaF$_2$ and PVT, because for these materials the high Compton to photo-electric cross section ratio means that no photopeak for Cs-137 may be observed.  We chose the FWHM energy resolutions of CaF$_2$ and PVT based on the energy resolution of NaI, and consideration of the number of optical photons produced by these materials relative to NaI (50\% and 25\% for CaF$_2$ and PVT, respectively).
Experiments with quite different geometries from ours have obtained energy resolutions
in PVT ranging from 10\%~\cite{tenperc_in_plastic} to 25\%~\cite{twentyfiveperc_in_plastic}. 
Of course, these quantities should be determined by experiment, for the particular configuration of scintillator and light-collection device chosen, and this will be the next phase of our detector design program.}.
To get an idea of the possible effect of underestimation of this parameter, we have simulated an additional PVT detector with a less optimistic FWHM energy resoluton of 25\% at 662~keV.
   \item A segmentation of the detector into 1~cm$^3$ pixels was also simulated.  When this effect is on, the energy deposit is assigned a reconstructed position at the centre of the pixel within which it took place.
\end{itemize}

In addition, attenuation of the signal in air has been simulated by including the air between the source and the detector in the simulation.

We also included an estimate of the effect of naturally occurring radioactive material (NORM).  To build up a spectrum for NORM, we followed a procedure similar to that outlined in~\cite{Novikova}.  The NORM energy spectrum had two components, a) ``lines'' - a set of lines representing the seven dominant energies emitted by the isotopes U-238, K-40, Th-232 and their daughters in equilibrium and b) ``continuum''- a continuum distribution which is domininant at low energies and monotonically decreasing with energy.  An isotropically emitting sheet of gamma rays with these raw spectra was passed through a simulation of a 
4~in~x~4~in~x~16~in NaI ``log'' and then compared with data taken with that log in a laboratory.  The comparison with the data was used to adjust the spectrum of the ``continuum'', and the ratio of ``lines'' to ``continuum'', until a reasonable representation of the low-energies and of the K-40, Bi-214 and Tl-208 peaks was obtained.  An isotropically emitting sheet source with that energy spectrum was then passed through the simulations presented here.

The rate of natural background which we have observed with this log in outdoor trials has ranged from $\sim$1000~s$^{-1}$ to $\sim$1600~s$^{-1}$.  To be conservative in our estimation of the amount of background in this study, we chose to allow that number of events to enter our model simulations which would correspond to a rate of 2000~s$^{-1}$ in our NaI log.  For 100,000 signal events, that comes to 293,460 NORM events.

\begin{figure*}[!t]
\centering
\includegraphics[width=6.5in]{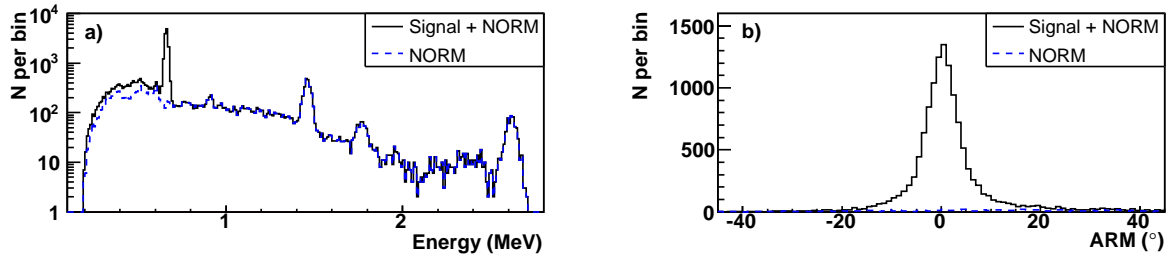}
\caption{For a 662~keV source at 20$^{\circ}$ off-axis and the NaI detector, for the golden event selection, a) shows the energy spectrum (leaving out the cuts on the peak energy) and b) shows the ARM distribution (including the requirement that the total energy falls within $\pm 3 \sigma$ of the peak energy).
The solid histogram shows signal and NORM.  The dashed histogram shows the contribution due to NORM separately.}
\label{fig_Spectrum_and_ARM}
\end{figure*}

For events with $> 50$~keV energy deposited in both the scatter and absorber detectors, and no more than one energy deposit in the scatter detector, we examined the spectra of the sum of the energies deposited in the scatter and absorber detectors.  The spectrum obtained with the NaI model for the 662~keV source at 20$^{\circ}$ off-axis is presented in Fig.~\ref{fig_Spectrum_and_ARM}~a).
A clear full-energy deposition peak was also observed for the other four
detector types, for all energies and angles.  Good fits to the peaks with the
sum of a Gaussian distribution and a straight line distribution were obtained.
The FWHM of the Gaussians fit to the full-energy deposition peaks
for the six different source energies, for an on-axis source, are presented in Table~\ref{etot_peaks} as a percentage.  (There is little dependence of this parameter on source angle.)

\begin{table}[!t]
\caption{FWHM (\%) of full-energy deposition peak (keV)}
\label{etot_peaks}
\centering
\begin{tabular}{|l|r|r|r|r|r|r|}
\hline
               &\multicolumn{6}{c|}{\bf Energy of source (keV)} \\
\cline{2-7}
{\bf Detector} & {\bf 300}& {\bf 500}&  {\bf 662}& {\bf 1000}& {\bf 1500 }& {\bf 2000} \\
\hline 
LaBr$_3$           &  2.9  &  3.7  &  3.2  &  2.4  &  1.9  &  1.7 \\
\hline
NaI                &  5.1  &  3.3  &  3.4  &  2.9  &  2.6  &  2.4 \\
\hline
CaF$_2$            &  7.7  &  6.2  &  5.1  &  4.3  &  3.6  &  3.3 \\
\hline
PVT                &        &      &       &       &       & \\
($\sigma_E =$14\%) &  9.6  &  7.2  &  7.0  &  4.6  &  4.6  & 4.1 \\
\hline
PVT                &        &      &       &       &       & \\
($\sigma_E =$25\%) & 12.3   & 12.3 & 10.1  &  7.4  &  7.1  & 6.6 \\
\hline
\end{tabular}
\end{table}

Note that the FWHM energy resolutions on the total energy, as presented in Table~\ref{etot_peaks}, are in some cases considerably better than one would expect from the smearing which has been applied to the energy deposits in the scatter detectors.
This is due to the fact that most of the energy is actually deposited in the absorber detector, so the overall detector energy resolution is dominated by the energy resolution of LaBr$_3$.

\section{Results}
\subsection{Cone Reconstruction}

We generated 100,000 signal events and 293,460 NORM events.  
``Golden'' events were defined as those which satisfy:
\begin{itemize}
   \item $> 50$~keV energy deposited in scatter detector,
   \item $> 50$~keV energy deposited in absorber detector,
   \item no more than one energy deposit in scatter detector, and
   \item total of energies deposited in scatter and absorber detector lying within three standard deviations of the maximum of the full-energy deposition peak.
\end{itemize}
\label{sec_golden}
For each golden event, we assigned the energy in the scatter detector to $E_{1}$ and the total energy in the absorber detector to $E_{2}$.  The position of the energy deposit in the absorber detector was assigned to the position of the maximum energy deposit there.  We then calculated the Compton cone opening angle according to Equation~\ref{eqn1}, and the angular resolution measure (ARM), the distance between the closest approach of the Compton cone back-projected onto a sphere, and the true source location. 

Figure~\ref{fig_Spectrum_and_ARM} b) shows an ARM distribution for a 662~keV point source 20$^{\circ}$ off-axis for the NaI model.
The ARM distribution is centered on zero degrees.  There is a small pedestal which is largely due to NORM.  Poorly reconstructructed signal events with escaping energy or misassignment of the first and second scatters also broaden the ARM distribution and contribute to the pedestal.  ARM distributions for the four other detector models look similar, with the models with poorer energy resolution exhibiting broader ARM distributions.

\begin{figure*}[!t]
\centering
\includegraphics[width=5.in]{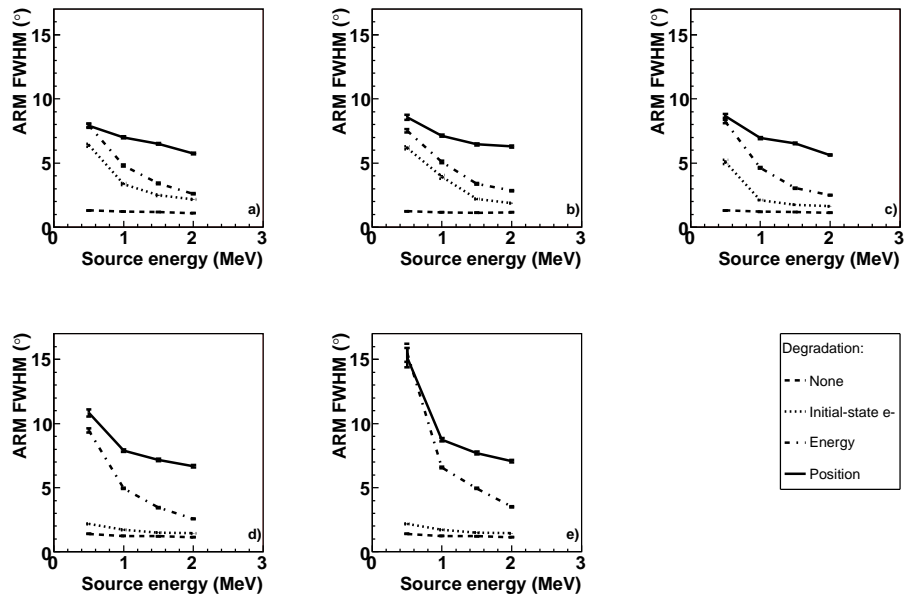}
\caption{FWHM ($^\circ$) of the ARM distributions for the a) LaBr$_3$, b) NaI, c) CaF$_2$,
d) PVT ($\sigma_E=14$\%) and e) PVT ($\sigma_E=25$\%) detectors as a function of source energy.
The dashed line shows the baseline image degradation due to event mis-reconstruction and background effects, the dotted line shows the effect of including initial state electron effects, the dash-dotted line shows the effect of including the energy resolution and finally, the solid line, shows the accumulation of all effects including the position segmentation of the detector.}
\label{fig_ARMS_vsE}
\end{figure*}

Figure~\ref{fig_ARMS_vsE} shows the FWHM of the ARM distributions for each detector model, for an on-axis point source, for four different source energies.
The degradation of image resolution due to initial-state electron effects, and the effects of energy resolution and position segmentation has been added successively to the simulation.
The dashed lines in Figure~\ref{fig_ARMS_vsE} show the result of allowing for effects such as back-scattering, which could lead to misassignment of the scatter and absorption occurrences.  No additional source of image degradation is included in the dashed curves.
The dotted curve shows the incremental image degradation due to including a detailed treatment of effects associated with the initial-state electron.
The dash-dotted curve shows the effect of including also the smearing of energies in the scatter and absorber detectors.
The solid curve is the final result for each detector.  It shows the ARM FWHM after every treated source of image degradation including segmentation of the scatter and absorber detectors into 1~cm$^3$ pixels.

This figure illustrates several well-known effects including, a) Doppler broadening affects high-Z materials more than low-Z materials and b) both Doppler broadening and energy resolution become less of a problem as source energy increases.  What is perhaps less well-known is the extent to which the 
smaller initial-state electron effects of the lower-Z materials can compensate for their worse
energy resolution.  The ARM FWHM values are comparable for all detector models at the higher energies. 

Figure~\ref{fig_ARMS_vsE} also illustrates that the
1~cm segmentation chosen for these simulations is a reasonable value
for these designs.  The additional image degradation observed by setting the
positions to the pixel centres is similar in magnitude to the other
detrimental effects.

\begin{figure*}[!t]
\centering
\includegraphics[width=5.in]{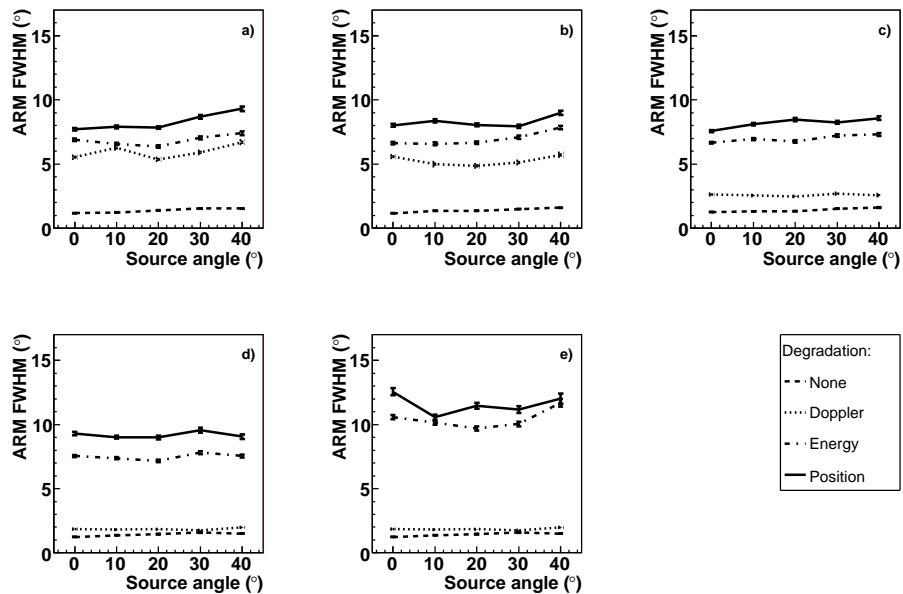}
\caption{FWHM of the ARM distributions for the a) LaBr$_3$, b) NaI, c) CaF$_2$,
d) PVT ($\sigma_E=14$\%) and e) PVT ($\sigma_E=25$\%) detectors as a function of source angle.
The dashed line shows the baseline image degradation due to event mis-reconstruction and background effects, the dotted line shows the effect of including initial state electron effects, the dash-dotted line shows the effect of including the energy resolution and finally, the solid line, shows the accumulation of all effects including the position segmentation of the detector.}
\label{fig_ARMS_vsTheta}
\end{figure*}

The FWHM of the ARM distributions are shown in Figure~\ref{fig_ARMS_vsTheta} for a 662~keV source at various angles with respect to the symmetry axis of the detector.

Our specifications call for an instrument with a wide field of view.  Fig.~\ref{fig_ARMS_vsTheta} illustrates that there is very little image degradation for the models studied here, out to angles of $\pm 40^{\circ}$.

Efficiency is defined as the percentage of those events which were generated in the direction of the front face of the detector, which satisfy the golden selection, and fall within three standard deviations of a Gaussian fit to the ARM distribution.

\begin{figure}[!t]
\centering
\includegraphics[width=3.in]{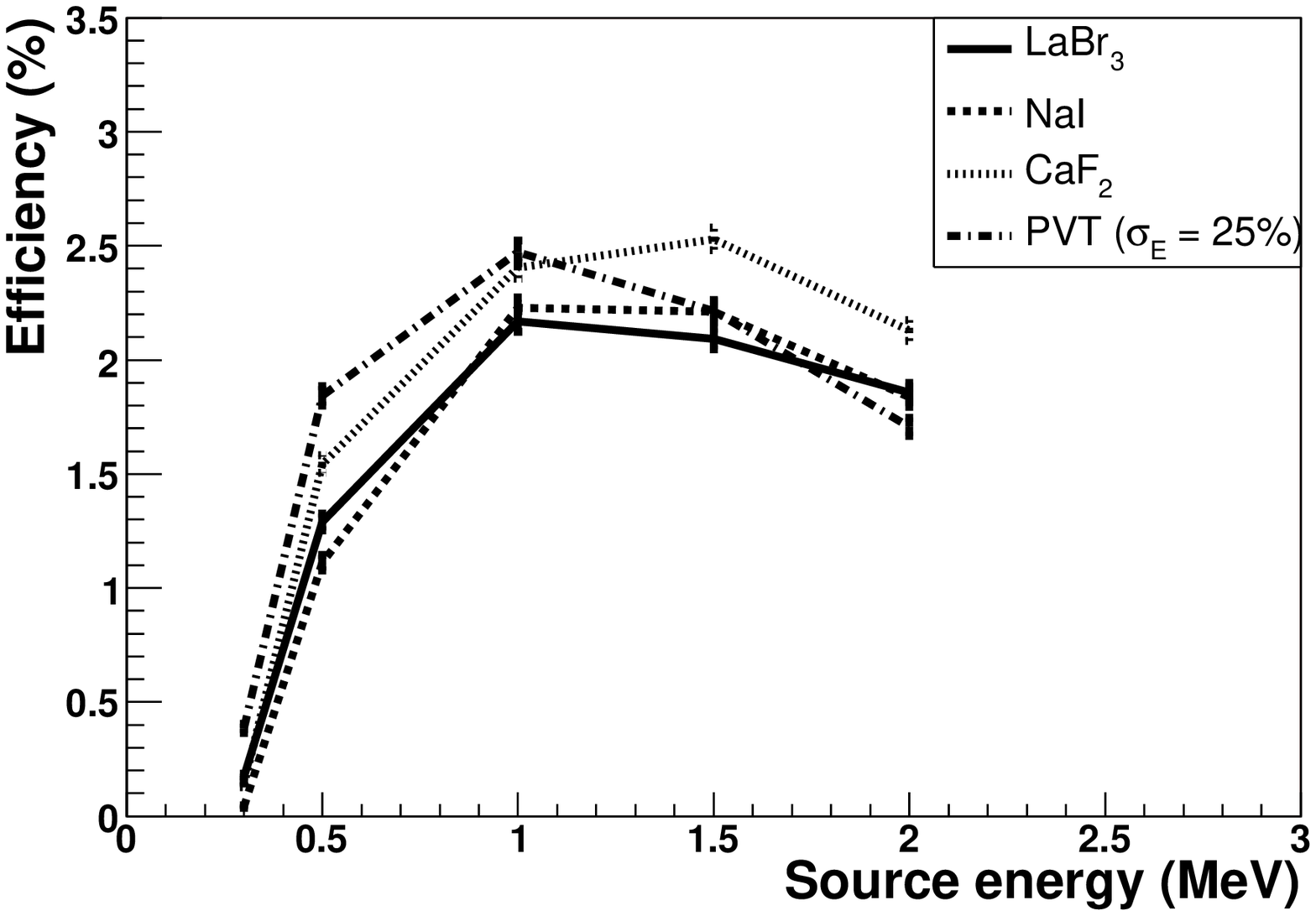}
\caption{Efficiency for varying source energy.
The solid, dashed, dotted and dash-dotted lines show the results for the 
LaBr$_3$, NaI, CaF$_2$, and PVT ($\sigma_E=25$\%) detectors respectively.
To reduce clutter in this plot the curve for PVT ($\sigma_E=14$\%) was left out.  It falls between those of CaF$_2$ and PVT ($\sigma_E=25$\%).}
\label{fig_EFF_vsE}
\end{figure}

\begin{figure}[!t]
\centering
\includegraphics[width=3.in]{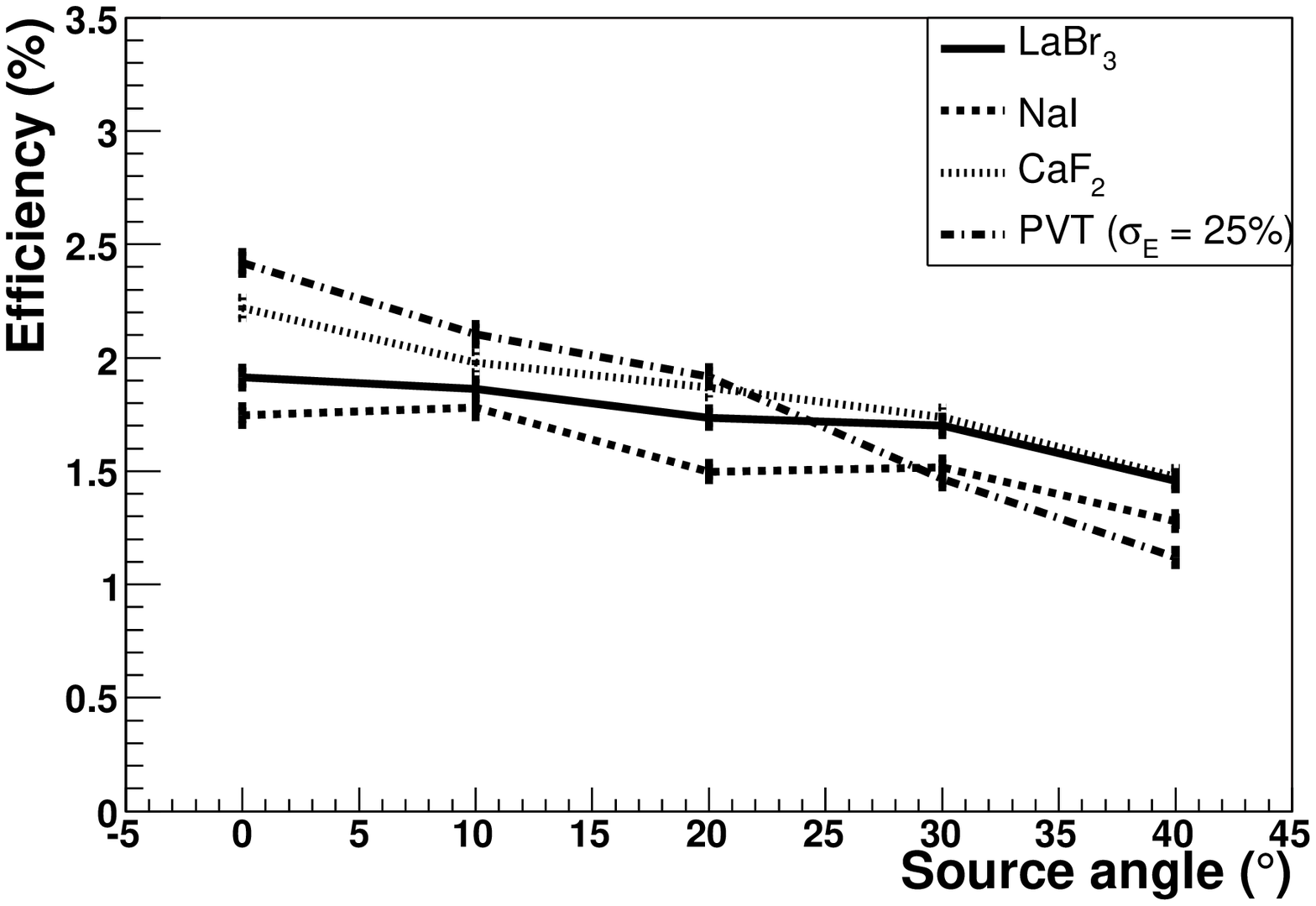}
\caption{Efficiency for varying angles between the line between the centre of the front face of the detector and the source, and the symmetry axis of the detector.
The solid, dashed, dotted and dash-dotted lines show the results for the 
LaBr$_3$, NaI, CaF$_2$, and PVT ($\sigma_E=25$\%) detectors respectively.
To reduce clutter in this plot the curve for PVT ($\sigma_E=14$\%) was left out.  This curve parallels that of PVT ($\sigma_E = 25$\%) but is 0.1 to 0.2~\% lower.}
\label{fig_EFF_vsTheta}
\end{figure}

Efficiencies are shown in Figures~\ref{fig_EFF_vsE} and~\ref{fig_EFF_vsTheta}, 
for an on-axis source of various energies and for a 662~keV source at various source angles, respectively.

In all of the studied models, the thickness of the scatter detector was chosen to optimize the acceptance of Compton events from Cs-137, which emits 662~keV gammas.  It is therefore not surprising that efficiency peaks at intermediate energies for all models.

Low efficiency at 300~keV prevents our consideration of these models for Compton gamma imaging for sources of energy below 500~keV.  A factor in the low efficiency is the lower energy threshold of 50~keV on the two energy deposits.  This value was chosen as an estimate of a practicable lower level threshold in the final design.  Variation of this threshold between 30 and 70~keV has no effect on our conclusions.  With care in the eventual instrument design, it may be possible to recover some events by lowering this threshold.

Another stringent requirement is the requirement of exactly one energy deposit in the scatter detector.  It should be possible in the future to improve efficiency by reconstructing more of the events which scatter at least twice anywhere in the scatter or absorber detectors.

Efficiency is highest for the on-axis source for the low-Z detectors.  The efficiency curve as a function of source angle is fairly flat for the detectors based on LaBr$_3$ and NaI, indicating that the field of view for those designs is good.  The PVT detector requires a lot of material in order to initiate the first Compton scatter and this geometric effect leads to the efficiency falling steeply toward the edge of the field of view.  With a more clever design for a detector based on PVT, where the absorber detector surrounds the scatter detector, the efficiency could be kept high.

Note that these studies based solely on the reconstruction of the Compton
scattering angles can give a lot of information about relative instrument
performance and dependencies on source characteristics, but they do not
directly answer the question of what image resolution can be expected in a
certain period of time.  That is the subject of the next section.

\subsection{Source Position Reconstruction}

As a final step in the analysis, the simulated data were passed through a 
position reconstruction procedure to determine the ability of each detector
to locate a point source. 

A standard \chsq\ minimization procedure has been employed to determine the direction vector
$\hat{v}(\theta,\phi)$ which best represents the direction from the centre of the front
face of the detector to the source.
For $N$ events, the following \chsq\ 
function was constructed with two fit parameters, $\theta$ and $\phi$, representing
the polar and azimuthal angles of the source direction vector to be determined:
\begin{equation}
\label{eq_chi2}
\chsq=\sum_{i=1}^{N}\left[ \frac{\mbox{ACA}(\theta,\phi)_i}{\sigma_{\mbox{\scriptsize ACA}i}} \right]^2
\end{equation}
where $\mbox{ACA}(\theta,\phi)_i$ is the angle of closest approach of the 
direction vector $\hat{v}$ to cone $i$, and $\sigma_{\mbox{\scriptsize ACA}i}$ is the uncertainty
on this angle. To simplify the $\mbox{ACA}$ expression, we have 
approximated the vertices of all cones as lying at the centre of the 
front face of the 
detector\footnote{This is an excellent approximation given a source at 40m from a detector of 
only 20~cm extent. For near sources, the adoption of a more accurate 
expression for $\mbox{ACA}$ would permit measurement of distance.}.  In this approximation, 
the expression for $\mbox{ACA}$ becomes: 
\begin{equation}
\label{eq_ACA}
\mbox{ACA} = |\mbox{acos } \hat{v}\cdot\hat{r_i}-\theta^C_i| 
\end{equation}
where $\hat{r_i}$ is the unit-vector axis for cone $i$.  Note that 
$\mbox{ACA}$ reduces to the absolute value of the ARM for the situation where $\hat{v}$ points
to the true source location.

The uncertainty on 
$\mbox{ACA}$, too cumbersome to reproduce here, is calculated on an event-by-event 
basis from the cone-axis uncertainty and the Compton opening angle uncertainty
(using Equation~\ref{eqn1}), taking into account the uncertainties on the energy
deposits and their positions. The effect of Doppler broadening is not 
included in the expression for the uncertainty. The \chsq\ expression was minimized using the 
Minuit package~\cite{Minuit}.   

To extract the source direction from a given sample of events, four fit 
iterations are performed, with the direction determined in each iteration 
passed on as a starting seed direction for the subsequent iteration.    
For the first iteration, a preliminary estimate of the source position is
deduced from the weighted mean of the cone axis directions, where 
for weights the squares of the opening angles are used. 
All events are included in the minimization procedure. In the second
iteration, events having a ratio $\mbox{ACA}/\sigma_{\mbox{\scriptsize ACA}}$ greater than five are 
excluded from participating in the fit. This step rejects  
background events and mis-identified or poorly reconstructed Compton events, 
but gives the fitting procedure ample freedom to find a new solution. In the 
third iteration, a rescan of the full set of $N$ events is performed, but 
only those events with $\mbox{ACA}/\sigma_{\mbox{\scriptsize ACA}}$ less than three are included in the 
fit. This provides further rejection of unwanted events, but permits events 
rejected from
the previous iteration, where a less accurate seed direction was used, to 
return to good standing. The final iteration is a repeat of the third, again
to take into account the more accurate starting seed vector. 

\begin{table*}
\caption{Average fit results (100 trials) for 3 seconds of acquisition time}
\label{table_fit}
\centering
\begin{tabular}{||l|c|c|c|c|c|c|c|c|c||}
\hline
\multicolumn{10}{|c|}{ Average fit results}\\
\hline
\hline
    &                 &                    &                  & NORM & NORM &       &  &  &  \\
Det & $\overline{N}$ & $\overline{N_{\rm fit}}$ & $\overline{\chi^2/{\rm dof}}$ & rejection & impurity & $\overline{\theta}$ & $\overline{\phi}$ & $\theta_{\rm RMS}$ & $\phi_{\rm RMS}$ \\
\hline
 PVT25 & 55 & 50 & 1.08 & 72\% & 10.8\% &   19.8$^\circ$ & -179.7$^\circ$ &    1.8$^\circ$ &    5.7$^\circ$ \\
\hline
 PVT14 & 51 & 45 & 1.04 & 56\% & 4.6\% &   20.0$^\circ$ & -180.0$^\circ$ &    1.1$^\circ$ &    4.1$^\circ$ \\
\hline
  CaF2 & 56 & 46 & 1.23 & 45\% & 3.3\% &   19.9$^\circ$ & -180.2$^\circ$ &    1.0$^\circ$ &    3.1$^\circ$ \\
\hline
   NaI & 47 & 38 & 1.45 & 29\% & 2.0\% &   20.0$^\circ$ & -179.8$^\circ$ &    1.0$^\circ$ &    3.2$^\circ$ \\
\hline
 LaBr3 & 52 & 40 & 1.55 & 25\% & 1.9\% &   20.1$^\circ$ & -179.9$^\circ$ &    0.9$^\circ$ &    2.8$^\circ$ \\
\hline
\end{tabular}
\end{table*}

This procedure was applied to three seconds of simulated data
from a single 662~keV point source positioned at $\theta=20^{\circ}$ and $\phi=-180^{\circ}$
for each of the modelled detectors.
The fit procedure was repeated for 100 trials.
Table~\ref{table_fit} shows the results.
The first seven colums show average values over the 100 trials.
The last two 
columns give the RMS spread of the fitted direction parameters. 
The starting number of events $N$ for each trial is dictated by the 
golden-event rate for the given detector, and $N_{\rm fit}$ is the number 
entering into the final iteration of the fit. The mean $\chi^2/{\rm dof}$ 
values are near unity for both PVT detectors, for which Doppler broadening 
is a negligible contribution. However, the $\chi^2/{\rm dof}$ values increase
for the other detectors reflecting the image degradation due to Doppler broadening.
The NORM rejection 
(impurity) column indicates the percentage of rejected (accepted) events that 
are background due to NORM. The mean values of $\theta$ and $\phi$ reflect an unbiased 
reconstruction of the source position.  Their RMS values characterize how well
the direction of the source can be localised. We may conclude that the 
fitting procedure accurately reproduces the source direction, and that there
is little difference between the non-PVT detector results. 

\begin{figure}[!t]
\centering
\includegraphics[width=3.in]{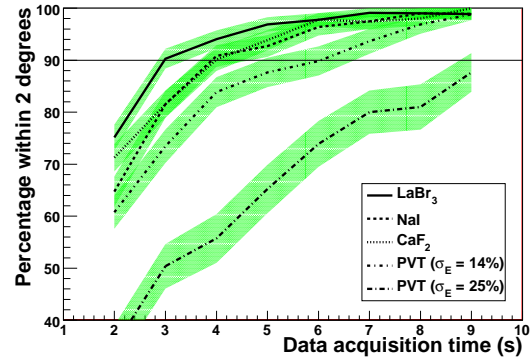}
\caption{Probability for correctly reconstructing the source direction to 
within $2^\circ$, as a function of data acquisition time. 
The solid, dashed, dotted, dash-dotted and long-dash dotted lines show respectively the results for the LaBr$_3$, NaI, CaF$_2$, PVT~($\sigma_E = 14\%$) and PVT~($\sigma_E = 25\%$) detectors.
The shaded band shows the statistical uncertainty.
The line at 90\% is 
to guide the eye.} 
\label{fig_prob}
\end{figure}

We also looked at the performance of the detector as a function of data acquisition
time.  For acquisition times of between two~seconds and nine~seconds we conducted between
500 and 100 fit trials.
Figure~\ref{fig_prob} shows the percentage
of fit trials for which the fitted direction was reconstructed to
within $2^\circ$ of the true source direction,
as a function of data acquisition time.
The 
bands show the uncertainty due to counting statistics. Except for 
PVT~($\sigma_E = 25\%$), 
all detectors are capable of correctly reconstructing the source direction 
most of the time even for small acquisition times. 
We find that LaBr$_3$ performs the best, with only three seconds of data required to
correctly reproduce the source location 90\% of the time.
CaF$_2$ and NaI 
are about the same with four seconds required for a 90\% success rate while
PVT~($\sigma_E = 14\%$) and PVT~($\sigma_E = 25\%$) are at six and ten 
seconds, respectively. These acquisition times are well under a minute, 
suggesting that any of the detector options we have considered is capable of 
meeting the design criteria.

\section{Conclusion}

The performance of four different all-scintillator Compton gamma imager models based on different materials for the scatter detector, has been investigated.
We have obtained encouraging results from all four of the scintillators looked at, 
LaBr$_3$, NaI, CaF$_2$, and PVT.
The all-LaBr$_3$ detector is predicted to perform the best of the models studied, 
with NaI and CaF$_2$ coming in a close second.
Indications are that even the worst model studied, with a scatter detector composed 
of PVT, may be able to provide an image of a 10~mCi 662~keV source at 40~m with an
RMS spread of the reconstructed source position of around two degrees, in under a minute.
The next stage of this work will be to establish a test stand to validate the Monte Carlo studies experimentally.

\section*{Acknowledgment}
The authors thank H.~Seywerd and J.~Carson for critical readings of the text.
This work will be proceeding to the development of a prototype, supported through funding from the Chemical, Biological, Radiological-Nuclear and Explosives, Research and Technology Initiative (CRTI Project 07-0193RD).

\ifCLASSOPTIONcaptionsoff
  \newpage
\fi


\end{document}